\documentclass[a4paper,fleqn,usenatbib]{mnras}

\usepackage{newtxtext,newtxmath}

\usepackage[T1]{fontenc}
\usepackage{ae,aecompl}

\usepackage{graphicx}	

\usepackage{amsmath}	
\usepackage{amssymb}	
\usepackage{ascmac}

\newcommand{\eg}{e.g., }

\def\gsim{\mathrel{\rlap{\lower 4pt \hbox{\hskip 1pt $\sim$}}\raise 1pt \hbox {$>$}}}
\def\lsim{\mathrel{\rlap{\lower 4pt \hbox{\hskip 1pt $\sim$}}\raise 1pt \hbox {$<$}}}

\newcommand{\Ea}{E_{\rm a}(\tau)}
\newcommand{\Ra}{R_{\rm area}(\tau)}
\newcommand{\Rv}{R_{\rm volume}(\tau)}

\title[A search for extragalactic FOT in the Tomo-e Gozen survey]{A search for extragalactic fast optical transients in the Tomo-e Gozen high-cadence survey} 

\author[K. Oshikiri et al.]
{Kakeru Oshikiri$^{1}$\thanks{E-mail: kakeru.oshikiri@astr.tohoku.ac.jp},
Masaomi Tanaka$^{1,2}$\thanks{E-mail: masaomi.tanaka@astr.tohoku.ac.jp},
Nozomu Tominaga$^{3,4,5}$,
Tomoki Morokuma$^{6}$,
\newauthor 
Ichiro Takahashi$^{7}$,
Yusuke Tampo$^{8}$,
Hamid Hamidani$^{1}$, 
Noriaki Arima$^{9}$,
Ko Arimatsu$^{10}$,
\newauthor
Toshihiro Kasuga$^{3}$,
Naoto Kobayashi$^{11}$,
Sohei Kondo$^{11}$,
Yuki Mori$^{11}$,
Yuu Niino$^{11}$,
\newauthor
Ryou Ohsawa$^{3}$,
Shin-ichiro Okumura$^{12}$,
Shigeyuki Sako$^{9,13,14}$,
and
Hidenori Takahashi$^{11}$
\\
\\
$^{1}$ Astronomical Institute, Tohoku University, Aoba, Sendai 980-8578, Japan\\
$^{2}$ Division for the Establishment of Frontier Sciences, Organization for Advanced Studies, Tohoku University, Sendai 980-8577, Japan\\
$^{3}$ National Astronomical Observatory of Japan, National Institutes of Natural Sciences, 2-21-1 Osawa, Mitaka, Tokyo 181-8588, Japan\\
$^{4}$ Department of Astronomical Science, School of Physical Sciences,\\
\ \ \ The Graduate University of Advanced Studies (SOKENDAI), 2-21-1 Osawa, Mitaka, Tokyo 181-8588, Japan\\
$^{5}$ Department of Physics, Faculty of Science and Engineering, Konan University, 8-9-1 Okamoto, Kobe, Hyogo 658-8501, Japan\\
$^{6}$ Planetary Exploration Research Center, Chiba Institute of Technology, 2-17-1 Tsudanuma, Narashino, Chiba 275-0016, Japan\\
$^{7}$ School of Science, Tokyo Institute of Technology, 2-12-1 Ookayama, Meguro-ku, Tokyo 152-8551, Japan\\
$^{8}$ Department of Astronomy, Kyoto University, Kitashirakawa-Oiwake-cho, Sakyo-ku, Kyoto, Kyoto 606-8502, Japan\\
$^{9}$ Institute of Astronomy, Graduate School of Science, The University of Tokyo, 2-21-1 Osawa, Mitaka, Tokyo 181-0015, Japan\\
$^{10}$Hakubi Center / Astronomical Observatory, Graduate School of Science, Kyoto University, Kitashirakawa-Oiwakecho, Sakyo-ku, Kyoto 606-8502, Japan\\
$^{11}$Kiso Observatory, Institute of Astronomy, Graduate School of Science, The University of Tokyo, 10762-30 Mitake, Kiso-machi, Kiso-gun, Nagano 397-0101, Japan\\
$^{12}$Japan Spaceguard Association, Bisei Spaceguard Center, 1716-3 Okura, Bisei, Ibara, Okayama 714-1411, Japan\\
$^{13}$UTokyo Organization for Planetary Space Science, The University of Tokyo, 7-3-1 Hongo, Bunkyo-ku, Tokyo 113-0033, Japan\\
$^{14}$Collaborative Research Organization for Space Science and Technology, The University of Tokyo, 7-3-1 Hongo, Bunkyo-ku, Tokyo 113-0033, Japan\\
}

\date{Accepted XXX. Received YYY; in original form ZZZ}

\pubyear{2023}

\begin{document}
\label{firstpage}
\pagerange{\pageref{firstpage}--\pageref{lastpage}}
\maketitle

\begin{abstract}
The population of optical transients evolving within a time-scale of a few hours or a day (so-called fast optical transients, FOTs) has recently been debated extensively. 
In particular, our understanding of extragalactic FOTs and their rates is limited. 
We present a search for extragalactic FOTs with the Tomo-e Gozen high-cadence survey. 
Using the data taken from 2019 August to 2022 June, we obtain 113 FOT candidates. 
Through light curve analysis and cross-matching with other survey data, 
we find that most of these candidates are in fact supernovae, variable quasars, and Galactic dwarf novae, that were partially observed around their peak brightness. 
We find no promising candidate of extragalactic FOTs. 
From this non-detection, we obtain upper limits on the event rate of extragalactic FOTs as a function of their time-scale. 
For a very luminous event (absolute magnitude $M<-26$ mag), we obtain the upper limits of 
$4.4\times 10^{-9}$ Mpc$^{-3}$ yr$^{-1}$ 
for a time-scale of 4 h, and 
$7.4\times 10^{-10}$ Mpc$^{-3}$ yr$^{-1}$ 
for a time-scale of 1 d. 
Thanks to our wide (although shallow) surveying strategy, 
our data are less affected by the cosmological effects, 
and thus, give one of the more stringent limits to the event rate of intrinsically luminous transients with a time-scale of $< 1$ d. 
\end{abstract}



\section{Introduction}
\label{sec:1}

Astronomical objects with rapid rise and decay in their luminosity are called transients. 
Transients with relatively long time-scales, of weeks to months, 
such as supernovae (SNe) and classical novae (CNe), 
have been intensively observed and studied; 
thanks to their relatively long time-scales and reasonably high event rates 
[$\sim 10^{-4}$ Mpc$^{-3}$ yr$^{-1}$ for SNe \citep{Li2011} and $\sim 30$ Galaxy$^{-1}$ yr$^{-1}$ for CNe \citep{Kawash2022}, respectively]. 
In contrast to these well-known long time-scale transients, 
short time-scale transients, with time-scale of a few hours to days 
(hereafter, denoted as fast optical transients, FOTs) 
are more difficult to detect as their detection requires high-cadence surveys. 
Such high-cadence surveys had not been intensively performed as higher cadence observations usually result in smaller surveying area or lower sensitivity (shallower limiting magnitude).

The development of wide-field instruments and improvements in computational power that enables automatic, high-speed processing of big data have enabled observers to monitor wide area with short time cadences and with deep limiting magnitudes. 
Thanks to these developments, our understanding of extragalactic transients with time-scales of a few days has greatly advanced. 
For example, the population of so-called fast blue optical transients (FBOTs; \eg \citealt{drout2014,Pursiainen2018,margutti2019}), such as AT 2018cow, has been newly discovered. 
FBOTs show a very high luminosity (absolute magnitude of $M \lsim -20$ mag) and show rapid brightness variations over several days. 
Although their origin is still under debate \citep[\eg][]{Prentice2018,Perley2019},
these objects have been observed not only in optical but also in radio, infrared, ultraviolet, and X-ray wavelengths. 
In addition, minute time-scale flare has recently been reported at the position of FBOT \citep{ho2022_tns}, which makes the origin of these FBOTs even more enigmatic. 
Another example of short time-scale transient is the so-called ``kilonova" (KN), such as AT 2017gfo that was discovered as an optical/infrared counterpart of gravitational wave event GW170817 \citep{Abbott2017GW}. 
KN is a transient caused by binary neutron star merger, 
and shows a lower luminosity and faster time evolution as compared to normal SNe \citep[\eg][]{Lipa1998,metzger2010,rosswog2005,metzger2010,metzger2012,barnes2013,tanaka2013,metzger2019}.

In contrast to the advances made in understanding transients with a time-scale of days, 
our understanding for transients with a time-scale of a few hours (hour time-scale FOTs) is still limited. 
Several predictions have been made for such short time-scale transients. 
\cite{Villar2017} show that adiabatic expanding explosion of low ejecta mass can produce transients with duration < 1 d and low luminosity (absolute magnitude $M \leq -12$ mag). 
Also, transients with relativistic jets, such as GRB afterglows, are expected to be observed as luminous FOTs \citep[\eg][]{vanparadijs2000,rossi2002}. 
Some tidal disruption events (TDEs) are also known to be associated with relativistic jets (jetted TDEs, \eg \citealt{Bloom2011,Levan2011}), which may produce similar luminous, fast transients as GRBs. 
However, there might still be unknown populations of transients in this time-scale, and thus, 
the discovery of hour time-scale transients would push the frontier of time-domain astronomy. 
Also, understanding the populations of FOTs is important for future multimessenger observations: such population of transients with a time-scale of $<$ day would contaminate transient surveys triggered by the detection of gravitational waves, which are intended to search for rapidly evolving KNe.

To explore hour time-scale FOTs, there have been several active searches in the last decades, including 
the Deep Lens Survey (DLS; \citealt{Becker2004}), 
a search with Master telescope (\citealt{Lipunov2007}), 
the sub-hour survey in the Fornax galaxy cluster (\citealt{Rau2008}), 
a search in the Pan-STARRS1 Medium Deep Survey (PS1/MDS; \citealt{Berger2013}), 
the survey for minute time-scale transients with the Dark Energy Camera (DECam; \citealt{Andreoni2020}), 
the Sky2Night program (S2N; \citealt{Roestel2019}) using the Palomar Transient Factory (PTF; \citealt{Law2009}), 
and a search with the intermediate PTF (iPTF; \citealt{Ho2018}). 
However, there are not many extragalactic FOTs firmly been identified within 
the time-scale of hours, except for optical afterglows of GRBs (orphan afterglow) discovered without $\gamma$-ray trigger \citep[\eg][]{Cenko2015,Andreoni2021,Ho2022} 
with PTF and Zwicky Transient Facility (ZTF, \citealt{Bellm2019}) 
and a luminous transient likely to be associated with a jetted TDE \citep{andreoni2022}. 
In fact, many of the objects found in the hour time-scale high-cadence transient surveys were of Galactic origin; 
notably stellar flares \citep[\eg][]{Becker2004,Berger2013,Ho2018,aizawa2022} and cataclysmic variables (CVs), in particular dwarf novae (DNe; \citealt{Rau2007}).

Searches for extragalactic FOTs and constraints on their event rate depend largely on the limiting magnitude, cadence, and area of the surveys. 
Therefore, independent searches with different observing strategies are important. 
In this paper, we present our search for extragalactic FOTs on time-scales of hours to days with the Tomo-e Gozen high-cadence survey. 
Our survey and selection process for candidates of FOTs are described in Section \ref{sec:2}.
Then, results are presented in Section \ref{sec:3}. 
Constraints on the event rate of extragalactic FOTs are discussed in Section \ref{sec:4}. 
Finally, a summary is presented in Section \ref{sec:5}. 
Throughout the paper, we assume the flat $\Lambda$CDM cosmological model ($H_0 = 67.7$ km s$^{-1}$, $\Omega _{\mathrm{m}} = 0.310$, $\Omega _{\mathrm{\Lambda}} = 0.69$; \citealt{planck2020}). 

\section{Observations and Data Analyses}
\label{sec:2}

\subsection{Tomo-e Gozen high-cadence survey}
\label{sec:tomoe} 

Our data were collected with the Tomo-e Gozen, a wide-field camera mounted on the 1.05 m Kiso Schmidt telescope \citep{Sako2018}. 
The camera consists of 84 CMOS image sensors: 
each sensor has 2000 $\times$ 1128 pixels with a pixel scale of 1.189 arcsec per pixel, covering 0.246 deg$^2$ area. 
A simultaneous field of view of the camera is 20.7 deg$^2$. 
All data used in this paper have been taken with no filter. 
Photoelectric conversion efficiency of the CMOS sensor peaks around 5000 \AA \
(with an efficiency of 0.72), and it declines to half of the peak at 3800 and 7100 \AA.

A high-cadence transient survey with Tomo-e Gozen has been performed since 2019 August. 
In this paper, we use the data taken until 2022 June. 
The survey is designed to visit about 3000 deg$^2$ with a cadence of about 3--4 times per night, which provides the cadence to detect transients lasting a few hours. 
The entire visible sky is also visited every usable night. 
Although the main focus of this paper is transients with a time-scale of a few hours, 
we systematically search for transients with a time-scale of $< 5$ d (see Section \ref{sec:selection}). 

All data were processed with the Tomo-e Gozen standard pipeline and transient pipeline in a real-time manner. 
Data were taken with a rate of 2 frames per second. 
Then, 12 or 18 consecutive images were stacked together. 
Data used in this paper are these stacked data with an effective exposure time of 6 or 9 s. 
In the standard pipeline, bias and dark subtraction and flat-field correction are performed. 
After the standard reduction, 
source detection is performed using SExtractor \citep{Bertin1996}. 
Flux calibration is performed with respect to the Pan-STARRS1 (PS1) catalogue. 
The Tomo-e Gozen bandpass covers a wider wavelength range than a single normal broad-band. 
The zero-point magnitude of the Tomo-e Gozen data is determined relative to the converted magnitudes of the PS1 $g$, $r$, and $i$-band magnitudes of objects detected in the Tomo-e Gozen data. 
The converted magnitude is the sum of the PS1 $g$-band magnitude and 6th-order polynomial of the PS1 $g-i$ color. 
The coefficients of the 6th-order polynomial are obtained by calculating the synthetic magnitudes for the bandpasses of the Tomo-e Gozen and PS1 by using the Pickles stellar spectral library \citep{pickles1998}. 
After source detection and flux calibration, 
limiting magnitude (5 $\sigma$) is evaluated based on the significance of the detected sources. 
A typical 5 $\sigma$ limiting magnitude is about 18 mag (the median of the limiting magnitudes for the data used in this paper is 17.9 mag).

To search for transient objects, 
image subtraction is performed with hotpants package \citep{Becker2015}. 
For the reference image used in the image subtraction, 
we use PS1 $r$-band data \citep{Waters2020,Flewelling2020}. 
Here we did not use the deep reference images from Tomo-e Gozen because deep reference images were not available at the beginning of the survey. Note that the current survey (as of 2023) uses the deep reference image from Tomo-e Gozen generated by using the three-year survey data. 

After the image subtraction, source detection is performed for the subtracted images. 
To avoid bogus detection, a deep-learning real/bogus classifier is applied for the detected sources \citep{Takahashi2022}. 
The performance of the classifier has been tested by using the real transients registered in the Transient Name Server (TNS) \footnote{https://www.wis-tns.org}. 
We adopt a threshold of the classifier to keep the true positive rate of 0.9, which achieves a false positive rate of 0.0002. 
Then, the sources classified as real sources are matched with known asteroids by using astcheck software\footnote{https://www.projectpluto.com/astcheck.htm}, 
and removed if matched. 
Finally, if a source is detected more than once at the same position (< 3 arcsec), 
we register the source as a transient candidate (top of Figure \ref{fig:flowchart}). 
For the detected transient candidates, we produce the light curve by performing forced photometry for all the available images of Tomo-e Gozen centred on the coordinate of the first detection coordinate. 
Note that our real/bogus classifier \citep{Takahashi2022} has been updated in 2021 May. 
Since we also use the data before this update, we reapply the new classifier in the selection process to reduce the bogus detection (see Section \ref{sec:selection}). 

\subsection{Selection of extragalactic FOT candidates}
\label{sec:selection}

\begin{figure}
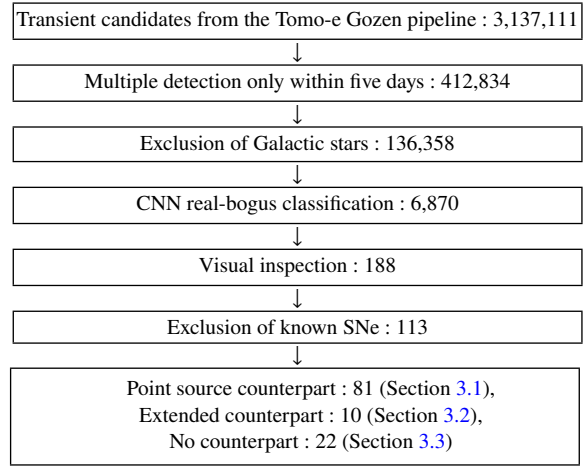

\begin{center}	
	\framebox[7.5cm][c]{Transient candidates from the Tomo-e Gozen pipeline : 3,137,111}\\
	$\downarrow$\\
	\framebox[7.5cm][c]{Multiple detection only within five days : 412,834}\\
	$\downarrow$\\
	\framebox[7.5cm][c]{Exclusion of Galactic stars : 136,358}\\
	$\downarrow$\\
	\framebox[7.5cm][c]{CNN real-bogus classification : 6,870}\\
	$\downarrow$\\
	\framebox[7.5cm][c]{Visual inspection : 188}\\
	$\downarrow$\\
	\framebox[7.5cm][c]{Exclusion of known SNe : 113}\\
 	$\downarrow$\\
    \framebox[7.5cm][c]{
    \begin{tabular}{c}
    Point source counterpart : 81 (Section \ref{sec:star}), \\
    Extended counterpart : 10 (Section \ref{sec:extragal}),\\ 
    No counterpart : 22 (Section \ref{sec:nc})
    \end{tabular}}	
    \caption{\label{fig:flowchart}
    Flowchart of the candidate selection process. 
    }
\end{center}
\end{figure}

For the transient candidates from the Tomo-e Gozen transient pipeline (see Section \ref{sec:tomoe}), 
we searched for extragalactic FOT candidates by the following steps (see Figure \ref{fig:flowchart}). 
First, we imposed multiple detections by the Tomo-e Gozen transient pipeline with a total duration of the detection within five days. 
Here, the detection duration is defined as the period between the first and the latest detection by the Tomo-e Gozen pipeline. 
In this way, we removed transients with a long time-scale, such as normal SNe and CNe, 
as well as variable stars and periodic transients that are expected to be detected for a longer duration. 

To focus on extragalactic transients, we removed objects associated with Galactic stars using Gaia Early Data Release 3 \citep[EDR3,][]{Gaia2021}. 
We considered Gaia objects that show significant ($>3\sigma$) parallaxes or proper motions as Galactic stars. 
Transient candidates located within 10 arcsec from the Gaia stars were excluded. 
By this criterion, transient associated with Galactic stars brighter than $\sim 20$ mag were removed. 
The remaining transient candidates are either transients associated with fainter Galactic stars or extragalactic transients. 
At this stage, there are still bogus detections of artefacts or imperfect subtracted images, in particular, for the transients detected in the period before the update of real/bogus classifier (see Section \ref{sec:tomoe}). 
Therefore, we reevaluated the candidates using a new deep learning classification with an improved performance \citep{Takahashi2022}. 

After these selection processes, 6870 objects remained. 
For these objects, we performed visual inspection. 
The majority of the excluded objects are false detections due to failed image subtraction. 
In this process, known minor planets, which Tomo-e Gozen pipeline failed to identify, were also removed by checking the Minor Planet Checker\footnote{https://cgi.minorplanetcenter.net/cgi-bin/checkmp.cgi}. 
Finally, we removed SNe which were already spectroscopically identified by cross-matching with the TNS. 
These SNe were included in our sample because they were detected only a few days near the peak brightness. 
As a result, we ended up with a total of 113 FOT candidates. 
In the next section, we discuss these candidates in detail.

\section{Results}
\label{sec:3}

\begin{figure}
  \begin{center}
    \includegraphics[keepaspectratio,width=\columnwidth]{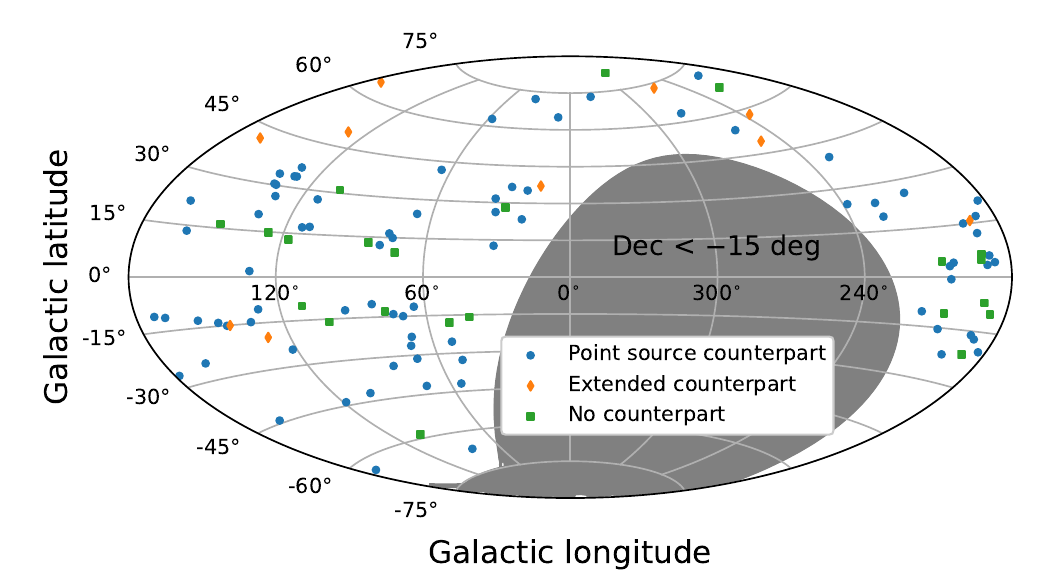}
\caption{
  \label{fig:map}
  The sky distribution of the final sample of 113 FOT candidates in the Galactic coordinate. 
  }
  \end{center}
\end{figure}

In this section, we classify our final sample of 113 candidates into three groups based on their association with the stacked object catalogue of the PS1 Data Release 2 \citep{Chambers2016, Flewelling2020}. 
The first group consists of objects associated with PS1 point sources that are likely to be faint Galactic stars, distant quasars, or distant galaxies. 
The second group consists of objects associated with extended sources that are likely to be galaxies. 
The third group consists of objects with no counterpart in the PS1 catalogue. 
Classification of the PS1 counterparts as point sources or extended sources 
is based on visual inspection of the PS1 images. 
Figure \ref{fig:map} shows the sky distribution of the final FOT candidates for each group in the Galactic coordinate. 
To obtain more information on our candidates, 
we also check ZTF data through ALeRCE \citep{sanchez2021}, TNS, 
NASA/IPAC Extragalactic Database (NED\footnote{https://ned.ipac.caltech.edu}), 
SIMBAD\footnote{http://simbad.cds.unistra.fr/simbad/}, and 
the International Variable Star Index (VSX\footnote{https://www.aavso.org/vsx/}, \citealt{Watson2006}).

Note that, for the light curves shown hereafter, we use the measured brightness by forced photometry in the Tomo-e Gozen images. 
If the measured flux exceeds $3 \sigma$, it is regarded as a detection. 
Usually, this gives more detection than automatic detection by Tomo-e Gozen transient pipeline ($\ge 5 \sigma$), and thus, 
light curves of some objects show detection longer than five days.

\subsection{Candidates with point source counterparts}
\label{sec:star}

\begin{figure}
  \begin{center}
    \includegraphics[keepaspectratio,width=\columnwidth]{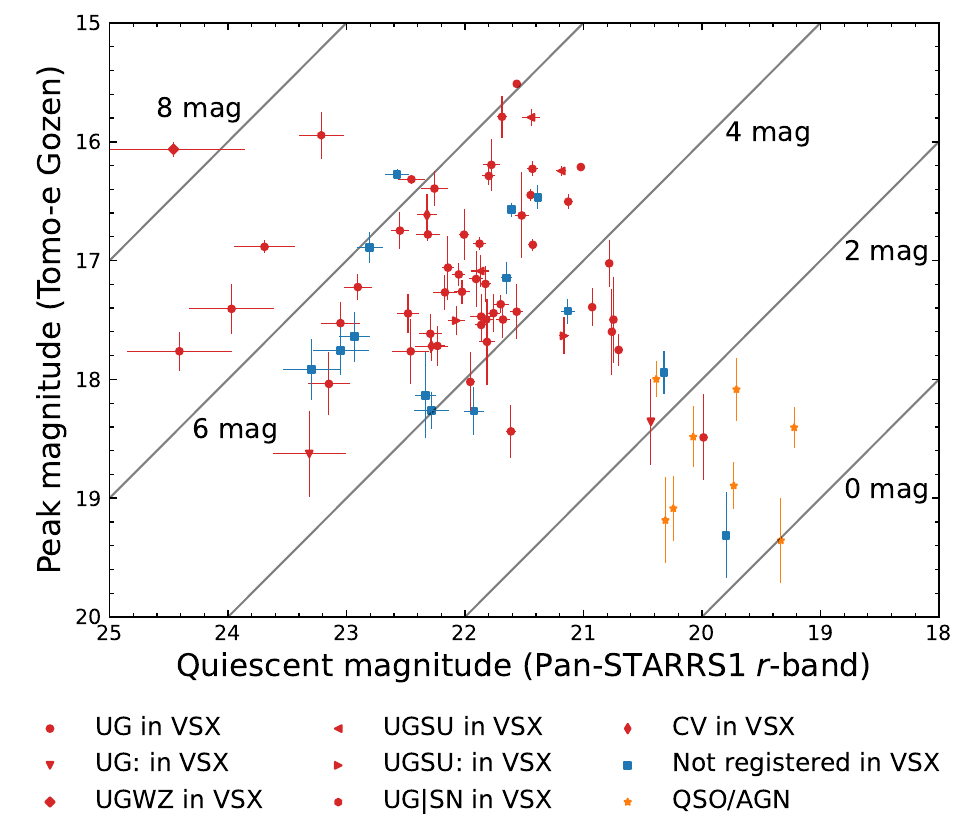}
   \caption{
  \label{fig:mag_ps1tomoe}
  Peak magnitude in Tomo-e Gozen versus quiescent magnitude in PS1 $r$-band for 81 candidates associated with PS1 point sources. 
  Candidates classified as QSO (quasar) or AGN (active galactic nucleus) in NED or ZTF are shown in orange. 
  Candidates registered in VSX are shown in red. 
  Here, UG stands for U Gem-type variables, often called dwarf novae. 
  UGWZ and UGSU are sub-types of UG. 
  The blue points are the other candidates which are not registered in VSX. 
  Each data point is plotted with a $1 \sigma$ error bar. 
  The grey solid lines represent outburst amplitude. 
  Note that the outburst amplitude is defined as difference between the PS1 quiescent magnitude in $r$-band minus the Tomo-e Gozen non-filter peak magnitude. 
  }
  \end{center}
\end{figure}

\begin{figure*}
  \begin{center}
    \includegraphics[keepaspectratio,width=1.8\columnwidth]{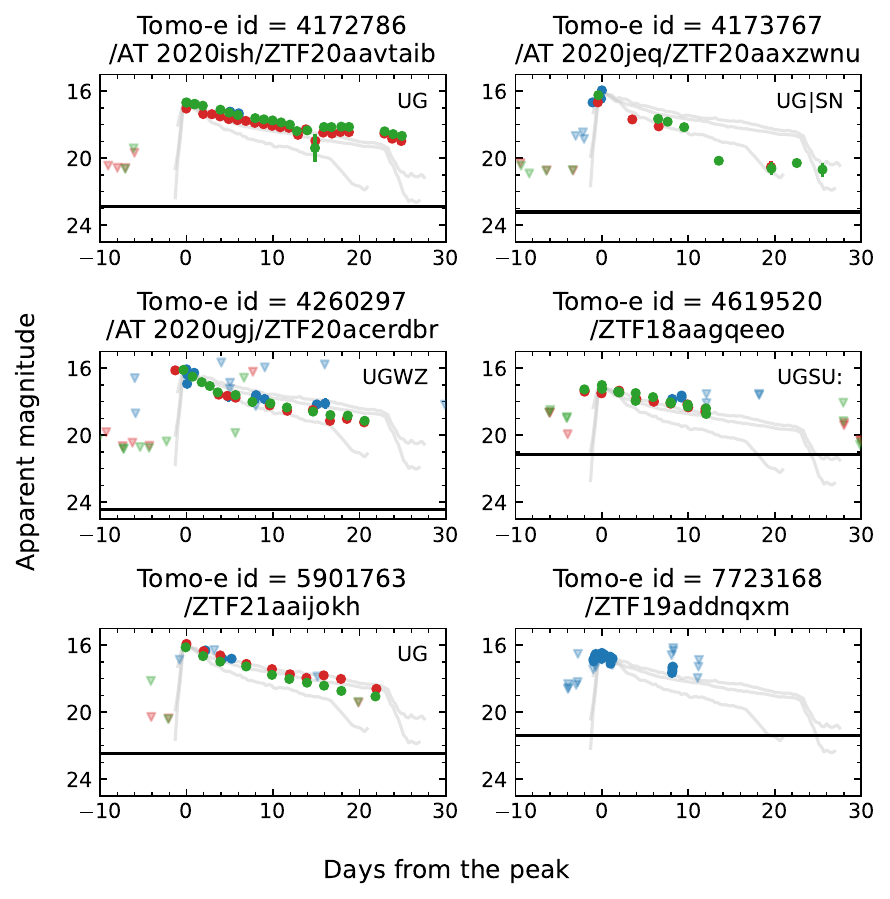}
   \caption{
  \label{fig:lc_star}
  Light curves of candidates associated with PS1 point sources. 
  The blue points show Tomo-e Gozen data while the red and green points show ZTF $r$-band and $g$-band data, respectively (the triangles are $3 \sigma$ upper limits). 
  The grey lines are light curves of WZ Sge-type DN outbursts (GW Lib,V455 And and WZ Sge, \citealt{kato2009}). 
  These light curves of DN outbursts are shifted to match the peak apparent magnitude of the FOT candidates. 
  The black horizontal line shows a quiescent magnitude in PS1 $r$-band. 
  }
  \end{center}
\end{figure*}

Among 113 candidates, 81 are associated with point sources in the PS1 catalogue, 
which are either objects fainter than Gaia sources or objects with insignificant parallax and proper motion. 
Thus, the objects in this group are likely to be Galactic transients 
or objects associated with distant quasars or galaxies. 
Figure \ref{fig:mag_ps1tomoe} shows peak magnitudes in Tomo-e Gozen for these 81 candidates as a function of the quiescent magnitude given in PS1 $r$-band. 
Gray solid lines represent amplitudes. 
We find that a majority of the objects (64 objects) have an amplitude of 2--6 mag. 
Eight candidates have > 6 mag amplitudes, including one exceeding 8 mag amplitude, 
while nine candidates have less than 2 mag amplitudes. 

For the Galactic transients with large amplitudes, DN outbursts are the most likely origins. 
DNe are binary systems that consist of a white dwarf and a low-mass star. 
The outburst is triggered by thermal instability in the accretion disk around the white dwarf \citep{osaki1996}. 
Such outbursts typically show 2--6 mag amplitude, lasting roughly a week \citep{Warner1995}. 
Moreover, so-called WZ Sge-type DNe show rare superoutburst with amplitudes reaching 9 mag \citep{kato2015}. 
For objects with a smaller amplitude, a stellar flare is a plausible candidate if the detection duration is within a few hours. 
Other possible candidates include a serendipitous detection of the brightest phase of variable stars or variable extragalactic objects such as quasars or blazars.

Since all of these transient/variable objects are known to show repeated activity, our candidates may have been discovered and classified in the past. 
Thus, we cross-matched our candidates with various catalogues. 
By searching VSX, 
we find that 59 candidates are registered and classified as DNe or CVs (red points in Figure \ref{fig:mag_ps1tomoe}). 
DNe, also known as U Gem-type variables, are represented as UG in Figure \ref{fig:mag_ps1tomoe}. 
WZ Sag-type and SU UMa-type variables, which are sub-types of UG, are denoted as UGWZ and UGSU, respectively. 
Furthermore, we also cross-matched our candidates with NED, SIMBAD, and ZTF ALeRCE light curve classifier. 
As a result, among 81 candidates, eight objects were classified as quasars or blazars (orange points in Figure \ref{fig:mag_ps1tomoe}). 
It is likely that they have been detected during an outburst only for a short duration (< 5 d).
The remaining 14 objects are not registered in either of VSX, NED, or SIMBAD. 
Of these, 13 objects show amplitudes consistent with DN outbursts (> 2 mag), 
while the other object shows < 2 mag amplitude (detected only with Tomo-e Gozen).

Figure \ref{fig:lc_star} shows the light curves of six representative objects in this group compared with those of known DN superoutbursts (WZ Sge-type). 
Combined with the deeper ZTF data, 
the light curves of many objects in this group show a good agreement with those of DN superoutbursts. 
In many cases, Tomo-e Gozen data capture the peak of the outburst only for a short period. 
As a result, these candidates satisfied our selection criteria.

In summary, based on the classification in VSX, NED, SIMBAD, and ZTF, 
59 objects in this group were found to be DNe and eight objects were identified as variable quasars/blazars. 
The other objects are most likely to be DN outbursts, 
considering the existence of associated point sources, amplitude and duration of the detection, and the properties of the light curves. 
Therefore, we conclude that there is no compelling evidence that the candidates in this group are extragalactic FOTs. 

\subsection{Candidates with extended counterparts}
\label{sec:extragal}

\begin{figure*}
  \begin{center}
    \includegraphics[keepaspectratio,width=1.8\columnwidth]{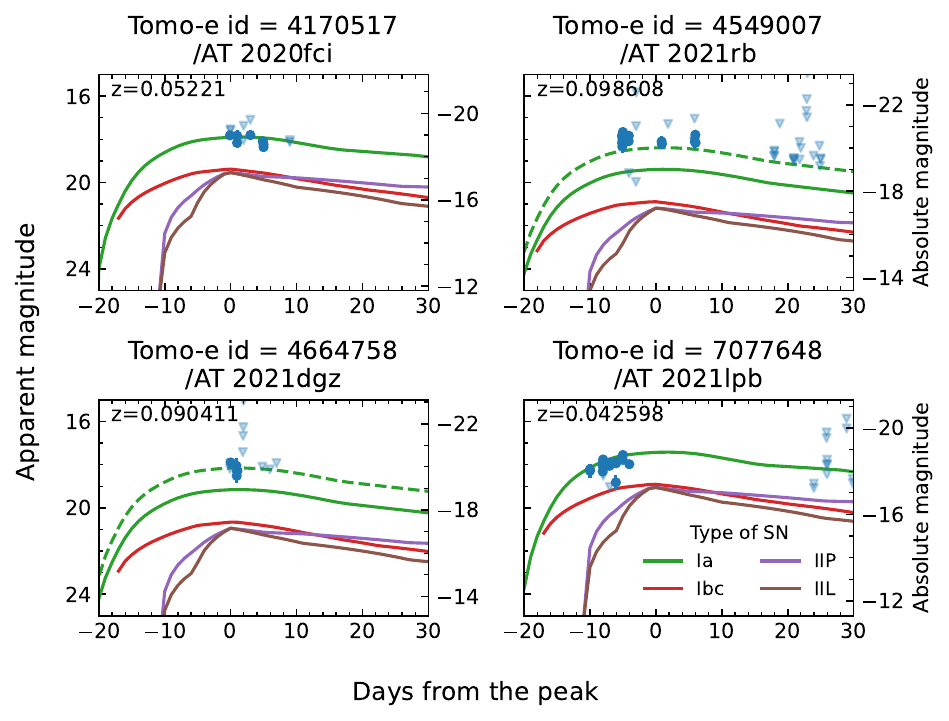}
   \caption{
  \label{fig:lc_g_z}
  Light curves of four FOT candidates that are associated with galaxies with known redshifts. 
  The blue points show Tomo-e Gozen data. 
  The green lines represent light curves of Type Ia SNe in SDSS $r$-band with peak absolute magnitude $M_{\mathrm{B}}$ = $-19$ mag (solid), $-20$ mag (dashed) and $-18$ mag (dotted). 
  Parameters of the SALT2 templates are fixed to be $x1$ = 0.945 and $c=-0.043$ as average properties of Type Ia SNe \citep{Scolnic2016}. 
  The other lines represent Type Ibc SN (red), Type IIP SN (purple) and Type IIL SN (brown) with peak absolute magnitude $M_{\mathrm{B}}$ = $-17$ mag. 
  }
  \end{center}
\end{figure*}

\begin{figure*}
  \begin{center}
    \includegraphics[keepaspectratio,width=1.8\columnwidth]{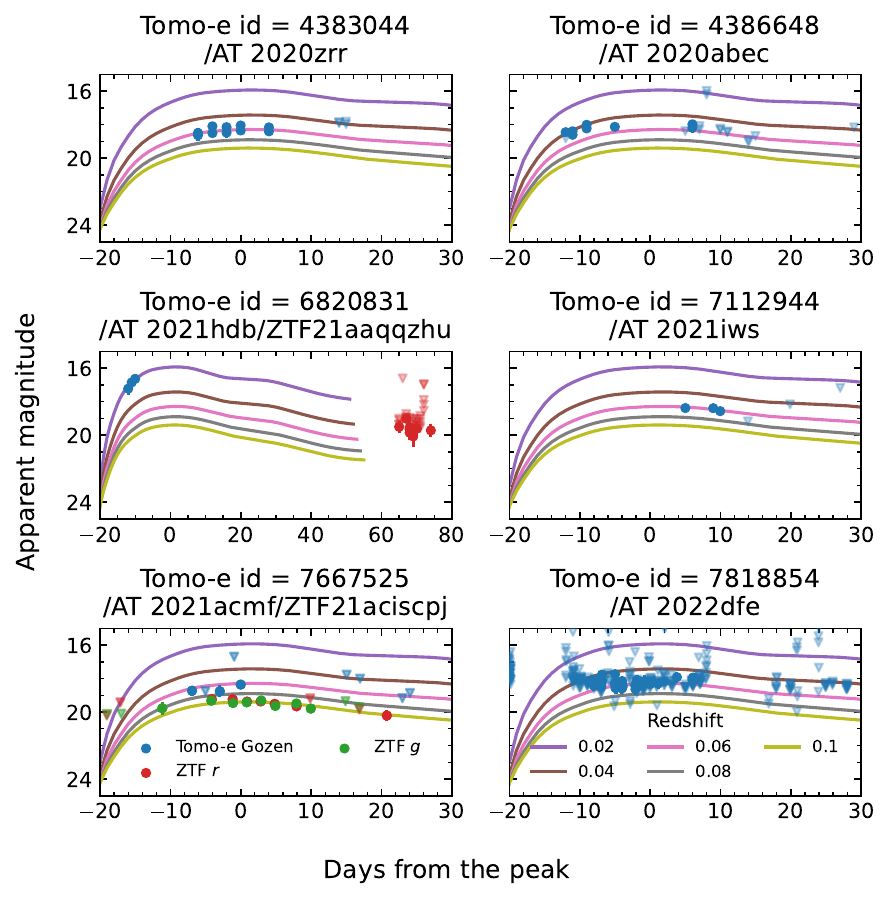}
   \caption{
  \label{fig:lc_g_nz}
   Light curves of six candidates that are associated with galaxies without redshift information. 
   The blue points show Tomo-e Gozen data while the red and green points show ZTF $r$-band and $g$-band data, respectively (the triangles are $3 \sigma$ upper limits). 
   The solid lines show the light curves of Type Ia SN models with a peak absolute magnitude $M_{\mathrm{B}}$ = $-19$ mag at $z$ = 0.02 (blue), 0.04 (yellow), 0.06 (green), 0.08 (red), and 0.10 (purple). 
  }
  \end{center}
\end{figure*}

Among 113 candidates, ten candidates are associated with extended counterparts in the PS1 catalogue. 
In fact, all of these sources are also detected by other transient surveys and are registered in TNS. 
However, their classification was not done due to the lack of spectroscopic follow-up observations. 
In Figures \ref{fig:lc_g_z} and \ref{fig:lc_g_nz}, we compare the light curves of these candidates with those of typical SNe using sncosmo package \citep{Barbary2016}. 
Here, we use SALT2 template \citep{Betoule2014} for Type Ia SNe, 
Nugent template for Type IIP, IIL, and Ibc SNe \citep{Gilliland1999,Levan2005}. 
 
Among ten candidates, four candidates are associated with galaxies whose redshift is given in the NED. 
Figure \ref{fig:lc_g_z} shows their light curves compared with those of SN templates at the same redshift. 
The comparison shows that the light curves of the four candidates are consistent with those of Type Ia SNe (green lines). 
Tomo-e Gozen captures the light curve only around the peak, 
and the evolution of the light curve for the detected period is not particularly fast.

Light curves of the other six candidates are shown in Figure \ref{fig:lc_g_nz}. 
For these objects, redshift of the host galaxies is not known, and thus, 
we only show the light curves of Type Ia SNe at some selected redshifts. 
As these objects are detected only for a few days, it is difficult to determine the type of SNe. 
Nevertheless, the light curves of these objects do not evolve very rapidly, 
and the light curves are consistent with those of known types of SNe around the peak. 
Therefore, we conclude that our ten candidates associated with extended objects in PS1 are not extragalactic FOTs.

\subsection{Candidates without counterparts}
\label{sec:nc}

\begin{figure*}
  \begin{center}
    \includegraphics[keepaspectratio,width=1.8\columnwidth]{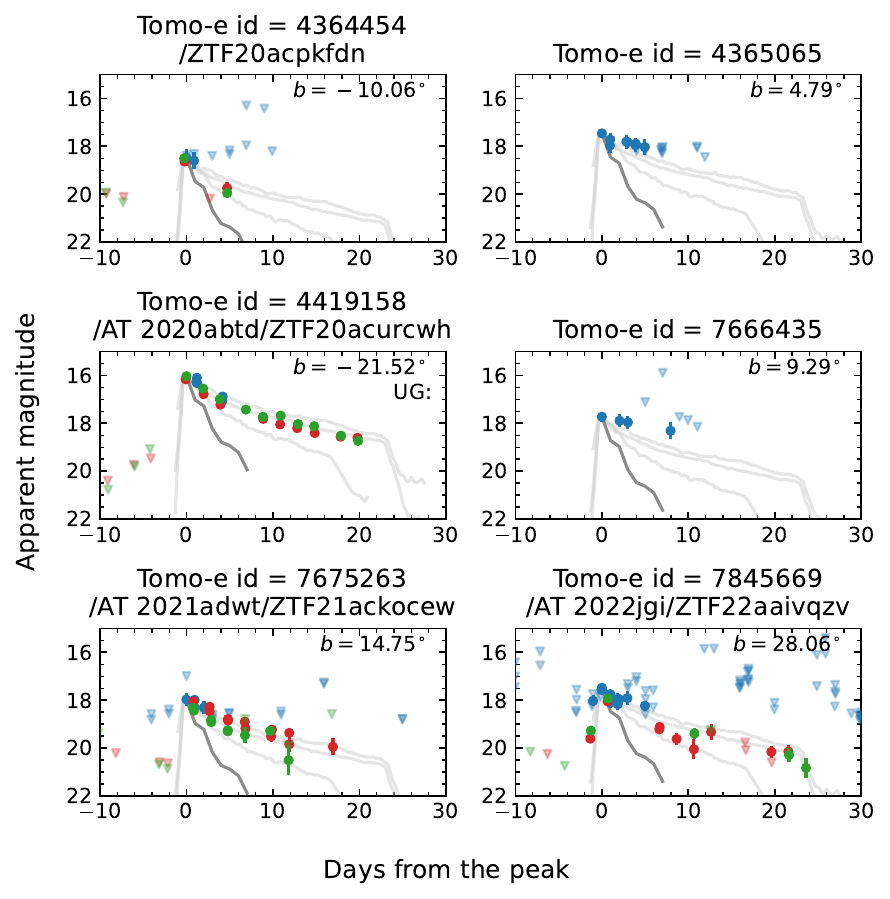}
   \caption{
  \label{fig:lc_nostar}
  Light curves of candidates without point source counterparts. 
  The blue points show Tomo-e Gozen data while the red and green points show ZTF $r$-band and $g$-band data, respectively (the triangles are $3 \sigma$ upper limits). 
  The gray lines are light curves of WZ Sge-type DN outbursts (GW Lib, V455 And and WZ Sge, \citealt{kato2009}). 
  The black line is a light curve of kilonova AT 2017gfo \citep{Cowperthwaite2017}. 
  These light curves are shifted to match the peak apparent magnitude of the FOT candidates. 
  $b$ indicates galactic latitude. 
  }
  \end{center}
\end{figure*}

\begin{figure}
  \begin{center}
    \includegraphics[keepaspectratio,width=1.05\columnwidth]{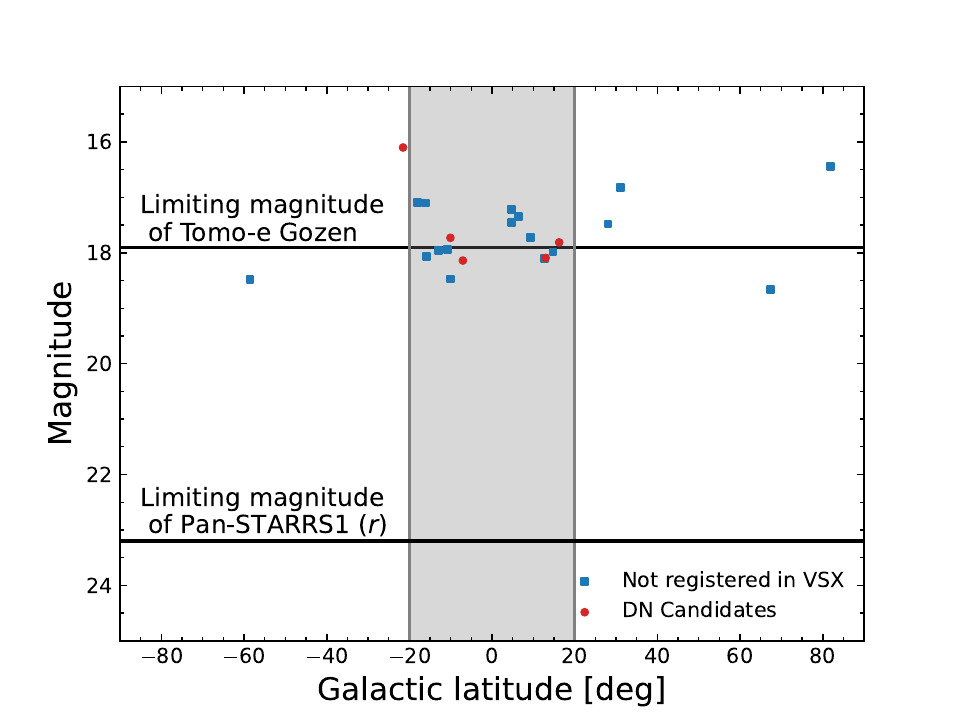}
   \caption{
  \label{fig:mag_nc}
  Peak magnitude in Tomo-e Gozen as a function of Galactic latitude for the candidates without PS1 counterparts. 
  Solid black lines represent typical limiting magnitudes of Tomo-e Gozen (17.9 mag) and PS1 $r$-band (23.2 mag). 
  DN candidates shown in red are objects classified as DN in VSX or the objects detected multiple outbursts in the past by ZTF or Gaia. 
  The grey shaded region indicates the low Galactic latitude $|b|<20^{\circ}$. 
 }
  \end{center}
\end{figure}

Twenty-two candidates are not associated with any counterparts in the PS1 catalogue. 
Figure \ref{fig:mag_nc} shows the peak magnitudes of these objects as a function of Galactic latitude. 
Solid lines represent typical limiting magnitudes of Tomo-e Gozen and PS1 $r$-band. 
As there is no detection in the PS1, their counterparts are fainter than the typical limiting magnitude of $r=$ 23.2 mag. 
We also confirmed that there is no counterpart in the GALEX ultraviolet catalogues \citep{Martin2005}. 
On the other hand, 13 objects are associated with infrared counterparts in NEOWISE \citep{Wright2010,Mainzer2011}. 
Their brightness and colour are similar to those of DN candidates with the PS1 counterparts shown in Section \ref{sec:star}.

Among these 22 candidates, three objects have been detected in the past by ZTF and Gaia; 
therefore, these three objects are most likely to be repeating sources such as high amplitude DNe. 
The remaining 19 objects are first time discoveries by Tomo-e Gozen, and they have not been spectroscopically observed so far. 
Thus, based on the information in the available catalogues alone, 
it is difficult to classify these transients as either extragalactic or Galactic objects. 

If these candidates are extragalactic objects, 
the lack of detection in the PS1 means that the difference of the brightness between the transients and host galaxies is > 5 mag. 
For example, for the case of Type Ia SNe with a peak absolute magnitude of about $-19$ mag, the host galaxy should be fainter than $-14$ mag in absolute magnitude. 
This is much fainter than the typical luminosity of the host galaxies of Type Ia SNe \citep[\eg][]{prieto2008,Childress2013,Perley2020}. 
The same is true for rapid transients like 
FBOT: their host galaxies are not necessarily faint \citep{wiseman2020}, and a $> 5$ mag contrast from the host galaxy is unlikely. 
On the other hand, for the case of superluminous supernovae (SLSNe), 
their peak absolute magnitude is about $-22$ mag and their host galaxies tend to be faint dwarf galaxies with about $-17$ mag \citep{Perley2016,Perley2020,avishay2019}; 
that is, their brightness contrast is large enough to be consistent with non-detection of the galaxy counterpart. 
However, photometric evolution of SLSNe are not necessarily fast \citep{avishay2019}. 
Thus, even if some of our candidates were SLSNe, we would not classify them as FOT candidates. 
In other words, there is no compelling evidence that our candidates are extragalactic FOTs.

Possible Galactic candidates for this group are DNe with > 5 amplitude magnitudes at a large distance where a quiescent source is not detectable in the PS1 images. 
Some WZ Sge-type DNe have a quiescent absolute magnitude of about 12 mag \citep[\eg][]{isogai2019,Abril2020}. 
Thus, if they are located at about 2.5 kpc, they are fainter than the apparent magnitude of 24 mag, which make them undetectable in the PS1 images. 
Hence, a hypothetical outburst with 6 magnitude amplitude would be observed with an apparent magnitude of 18 mag, and it can be recognised as a transient object without a counterpart. 

The sky distribution of the objects in this group is shown with green dots in Figure \ref{fig:map}. 
Among 22 objects, 16 objects are distributed at low Galactic latitude $|b|<20^{\circ}$. 
This fact may imply Galactic origin of these objects. 
If we assume a distance as expected for DN outbursts ($\sim$ 2.5 kpc), these objects would be located within $\sim 900$ pc height from the Galactic disc.

To further explore the nature of the objects in this group, 
we compare their light curves with those of some DN superoutbursts and KN AT 2017gfo \citep{Cowperthwaite2017}.  
The light curves of six objects are shown in Figure \ref{fig:lc_nostar} as representative examples. 
In fact, the light curves of these candidates are similar to those of WZ Sge-type DN superoutbursts, 
as in the cases for the objects with the counterpart (Section \ref{sec:star}).

To assess the possibility of the DN origin, we approximately compare our detection rate with the expected event rate of DNe. 
\cite{Rau2007} estimated the number of DN systems to be $(3.5--100) \times 10^3$ at $|b| < 45^{\circ}$ within 2 kpc assuming a range of the local number density of DN systems from $\sim 3 \times 10^{-5} \ \mathrm{pc}^{-3}$ \citep{Schwope2002} to $10^{-3}$ pc$^{-3}$ \citep{Kolb1993,deKool1993}. 
Assuming a mean outburst cycle of 1 yr, the yearly event rate of the outburst within this volume is $(3.5--100) \times 10^3$ yr$^{-1}$ at $|b| < 45^{\circ}$. 
Our Tomo-e Gozen survey has monitored about 25 per cent of the sky for about 3 yr. 
As a result, even by assuming about 50 per cent observing efficiency including weather factor, the number of the DN outburst in our survey data can be about $(1.3--40) \times 10^3$. 
It should be emphasised that our selection process removes the transients with bright counterparts in Gaia catalogue. 
Thus, the majority of nearby DN outbursts has been removed in our selection process, 
and thus, the number of DNe in our final candidates should be significantly smaller than in the above estimate. 
Nevertheless, given the high expected event rate, 
an order of 100 detections of DN outbursts in our samples (including those with PS1 counterparts) is not surprising. 
This is also true for the objects at a high Galactic latitude ($|b| > 45^{\circ}$). 

In summary, considering the spatial distribution, available light curve information, and the expected high event rate, 
we conclude that the majority of the candidates without PS1 counterparts can be explained by Galactic DN outbursts (at > 2.5 kpc). 

\section{Upper limits of the event rate of extragalactic FOTs}
\label{sec:4}

\begin{figure}
  \begin{center}
    \includegraphics[keepaspectratio,width=\columnwidth]{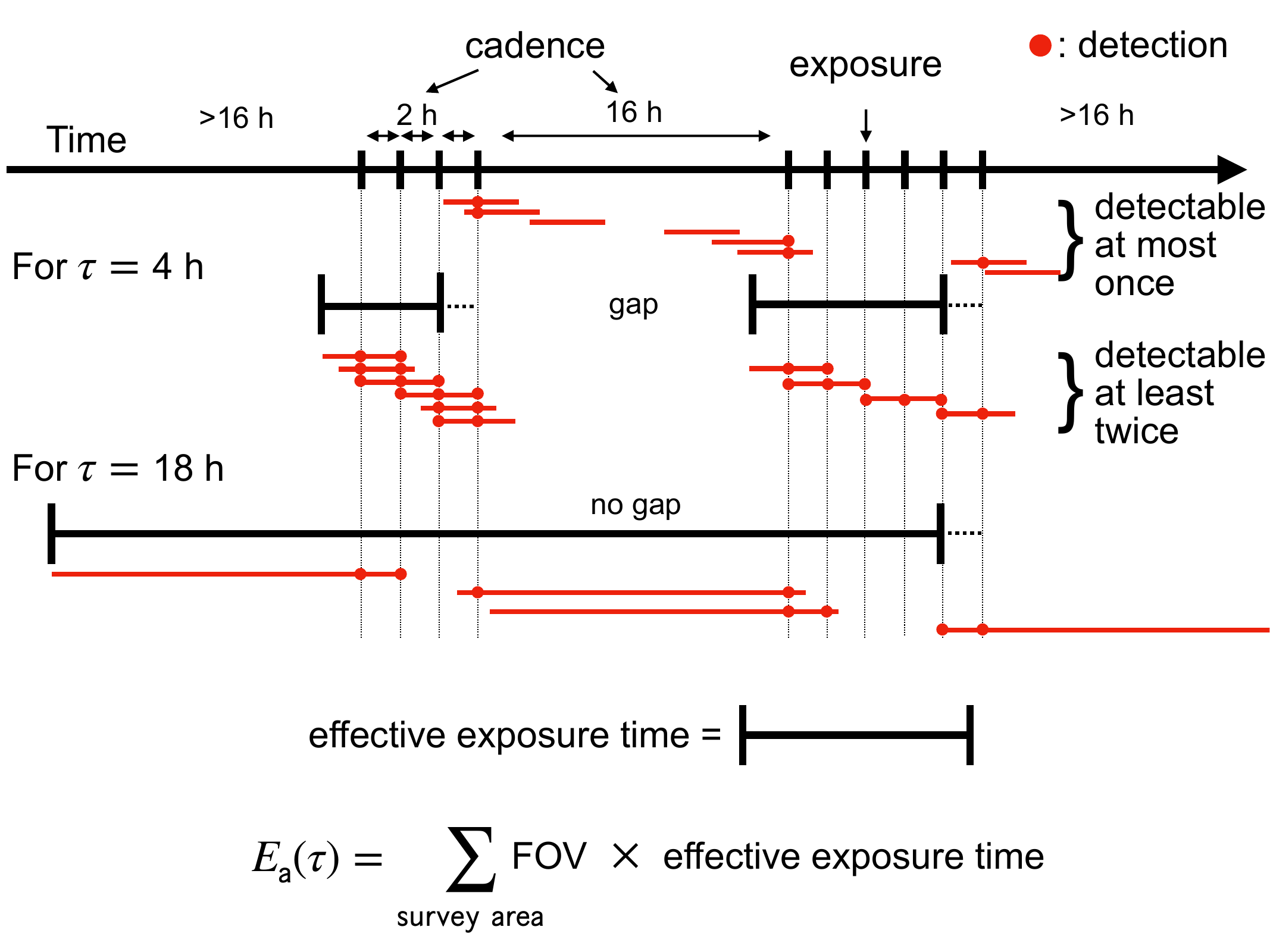}
   \caption{
  \label{fig:cal_Ea}
  A schematic explanation of the effective exposure time and areal exposure. 
  Two examples with different visible time-scales ($\tau=4$ and $18$ h respectively) are illustrated. 
  The effective exposure time is defined as the time when an object can be observed at least twice with for given visible time-scale $\tau$. 
  Transients occurring within the time interval shown by the black solid line segments can be detected at least twice, 
  while those occurring within the time interval shown by dotted horizontal lines can be detected only once. 
  The transients occurring within the gap in the black lines cannot be detected. 
  Hence, only the time interval represented by the black solid line segments are counted as effective exposure time. 
  When the visible time-scale $\tau$ is long enough, multiple detection over two nights become possible, which fills the gap. 
  The total areal exposure $\Ea$\ is calculated by summing up the areal exposure for each patch of the sky observed by our survey. 
  We followed the same method as \citet{Roestel2019}. 
  }
  \end{center}
\end{figure}

\begin{figure*}
    \begin{tabular}{cc}
      \begin{minipage}[t]{0.5\hsize}
        \centering
        \includegraphics[keepaspectratio, width=\columnwidth]{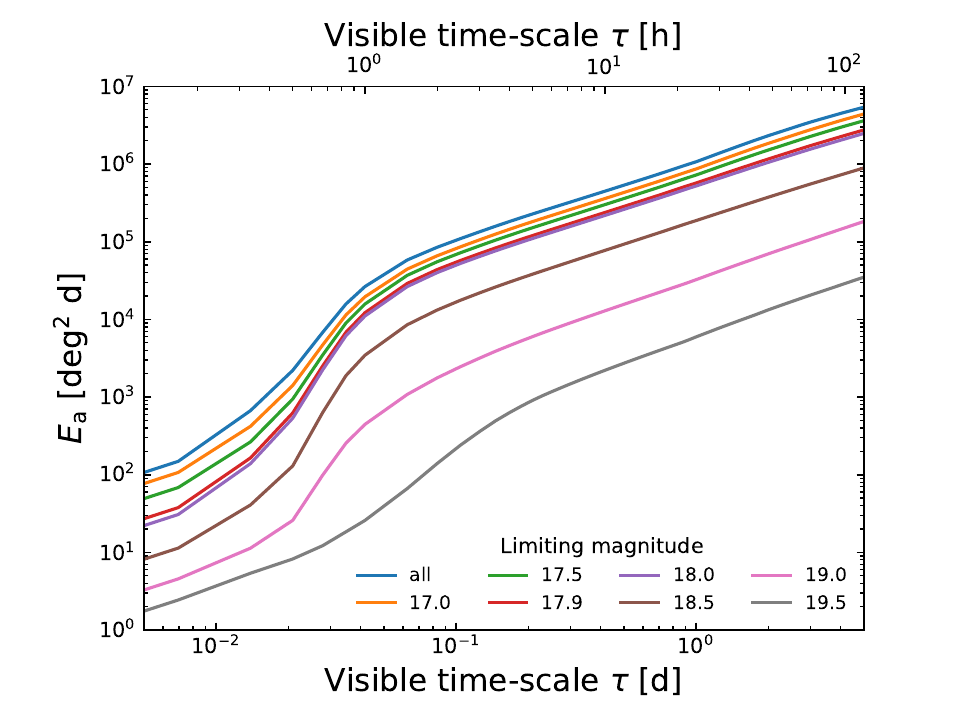}
        \label{fig:Ea}
      \end{minipage} &
      \begin{minipage}[t]{0.5\hsize}
        \centering
        \includegraphics[keepaspectratio, width=\columnwidth]{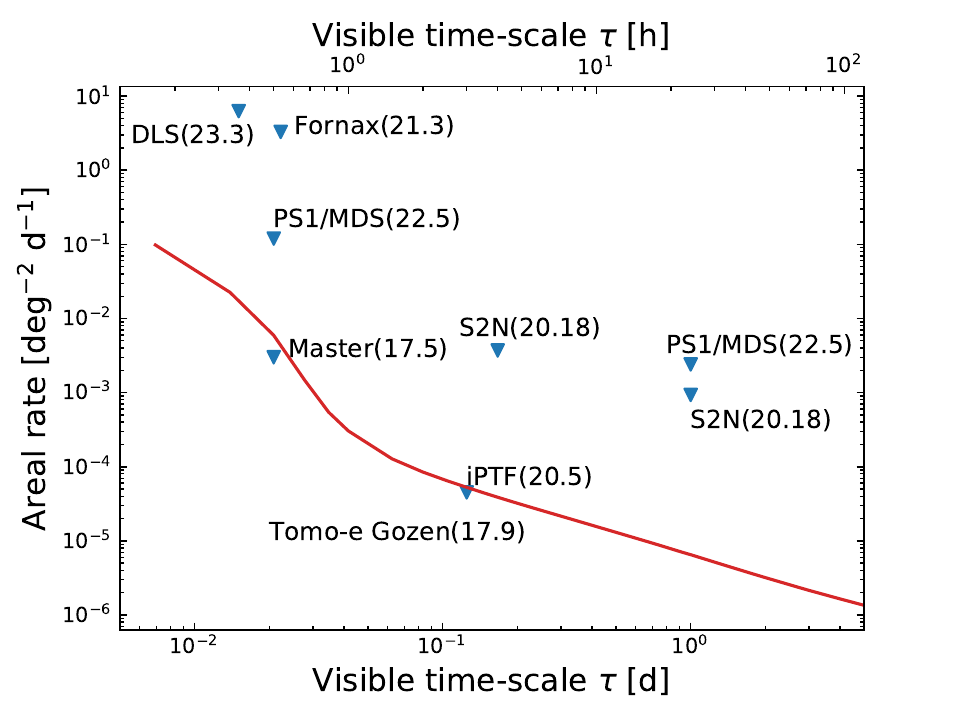}
        \label{fig:Ra}
      \end{minipage} 
      \\
      \begin{minipage}[t]{0.5\hsize}
        \centering
        \includegraphics[keepaspectratio, width=1.9\columnwidth]{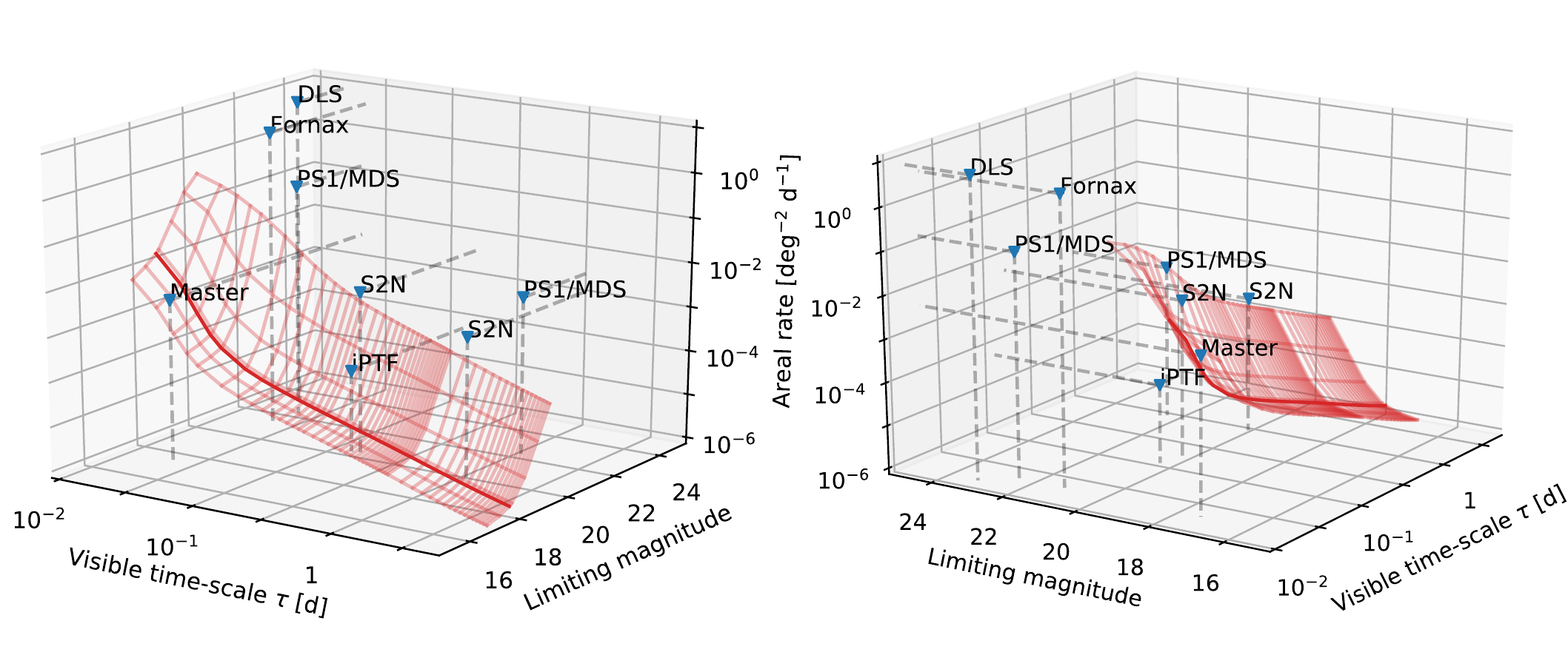}
        \label{fig:Ra_3D}
      \end{minipage}&
      \begin{minipage}[t]{0.1\hsize}
      \label{nothing}
      \end{minipage}
    \end{tabular}
     \caption{
     Top left: Areal exposure as function of visible time-scale $\tau$. 
     Different lines represent the cases with different limiting magnitudes. 
     Top right: Projection of the figures in the bottom panels along the axis of limiting magnitude to the plane of the areal rate and visible time-scale. 
     Bottom: Upper limit of the areal rate of extragalactic FOTs as function of visible time-scale and survey limiting magnitude. 
     The red surface represents the upper limit of the areal rate from our survey. 
     The dark red line indicates a median limiting magnitude of 17.9 mag. 
     The constraints from the Tomo-e Gozen survey are compared with those of other surveys: 
     DLS \citep{Becker2004}, Fornax \citep{Rau2008}, PS1/MDS \citep{Berger2013}, Master \citep{Lipunov2007}, iPTF \citep{Ho2018}, and Sky2Night \citep[S2N,][]{Roestel2019}.
     }
     \label{fig:EaRa}
  \end{figure*}

In this section, we discuss the constraints on the event rate of extragalactic FOTs. 
As discussed in Section \ref{sec:3}, we found no compelling candidates of extragalactic FOTs in our 113 final candidates: 
they are rather consistent with SNe, quasars/blazars, and outburst of Galactic DNe. 
Thus, we consider that the number of detected extragalactic FOTs is zero.

To evaluate the event rate, we first estimate the areal exposure 
$\Ea$\ [deg$^{2}$ d] of our survey, which is a measure of how wide of an area have been monitored with our survey and for how long. 
The areal exposure is expressed as the product of the survey area and effective exposure time which is sensitive to detect transients. 
Thus, the effective exposure time depends on the time-scale during which transient objects are visible (hereafter, this visible time-scale is identified as $\tau$). 
Figure \ref{fig:cal_Ea} schematically shows how the effective exposure time is calculated. 
Our definition of the effective exposure follows that by \citet{Roestel2019}. 
If a transient object with a visible time-scale $\tau$ appears within the time represented by the black solid line segments, 
it can be observed at least twice (as in our selection criterion, see Section \ref{sec:selection}). 
Then, the total length of the black line segment gives the effective exposure time. 

The areal exposure as a function of visible time-scale $\tau$ is shown in the top left panel of Figure \ref{fig:EaRa}. 
The slope of this function is the steepest at $\tau \simeq$ 0.04 d ($\sim$ 1 h) which corresponds to the typical cadence of the Tomo-e Gozen survey. 
Different lines represent the cases with different limiting magnitudes: if we limit the data with a deeper limiting magnitude, the areal exposure becomes smaller. 
To estimate the upper limit of the event rate below, 
we use images deeper than the mean $5 \sigma$ limiting magnitude of 17.9 mag, which is the median depth of the survey. 
Note that, 
we exclude Galactic stars in our selection process (see Section \ref{sec:selection}). 
Considering the number and distribution of Galactic stars in Gaia, 
there are about 6,000 stars per square degree in the region of $b \simeq 30^{\circ}$, 
corresponding to an unobserved area of approximately 0.15 deg$^2$ per 1 deg$^2$. 
In other words, about 15 per cent of the sky is not included in the survey. 
By taking 15 per cent as an average, this effect is taken into account in $\Ea$ shown in the top left panel of Figure \ref{fig:EaRa} and in the rate estimates below.

By using the areal exposure, we calculate an areal rate of extragalactic FOTs, $\Ra$. 
The areal rate is the number of transients per sky area per time: 
\begin{equation}
\Ra 
\ [\mathrm{deg}^{-2} \ \mathrm{d}^{- 1}] 
= \frac{N}{\epsilon^2 \ \Ea \ [\mathrm{deg^2 \ d}]},
\end{equation}
where $N$ is the number of detected transients and 
$\epsilon$\ is the detection efficiency per image.
Here, to give the upper limit of the event rate, we adopt $N \simeq 3$ (Poisson statistics with 95 per cent confidence). 
Since our real-bogus classifier gives a true positive rate of 0.9, we also assume $\epsilon = 0.9$ (see Section \ref{sec:tomoe}). 
The rate has a dependence of $\epsilon^2$ as we impose two detection.

The bottom panels of Figure \ref{fig:EaRa} show the comparison of the upper limit of the areal rate for extragalactic FOTs by different surveys as a function of visible time-scale and limiting magnitude. 
The top right panel in Figure \ref{fig:EaRa} shows the projection of the bottom panels along the axis of limiting magnitude to the plane of the areal rate and visible time-scale. 
This projected figure shows that Tomo-e Gozen gives a better limit to the areal rate than the other surveys at almost all the time-scale. 
For example, for a visible time-scale of $\tau \simeq 1$ h, 
the upper limit of the areal rate is $\Ra < 3.0 \times 10^{-4}$ deg$^{-2}$ d$^{-1}$. 
However, a simple comparison of $\Ra$ in this plane is not fair as the limiting magnitudes of the surveys are different. 
For example, our Tomo-e Gozen survey has a large $\Ea$ because of its wide survey area and short exposure time, while it is not necessarily deeper than other surveys. 
On the other hand, DLS and PS1/MDS are deeper but cover a relatively small survey area.

\begin{figure*}
  \begin{center}
    \includegraphics[keepaspectratio]{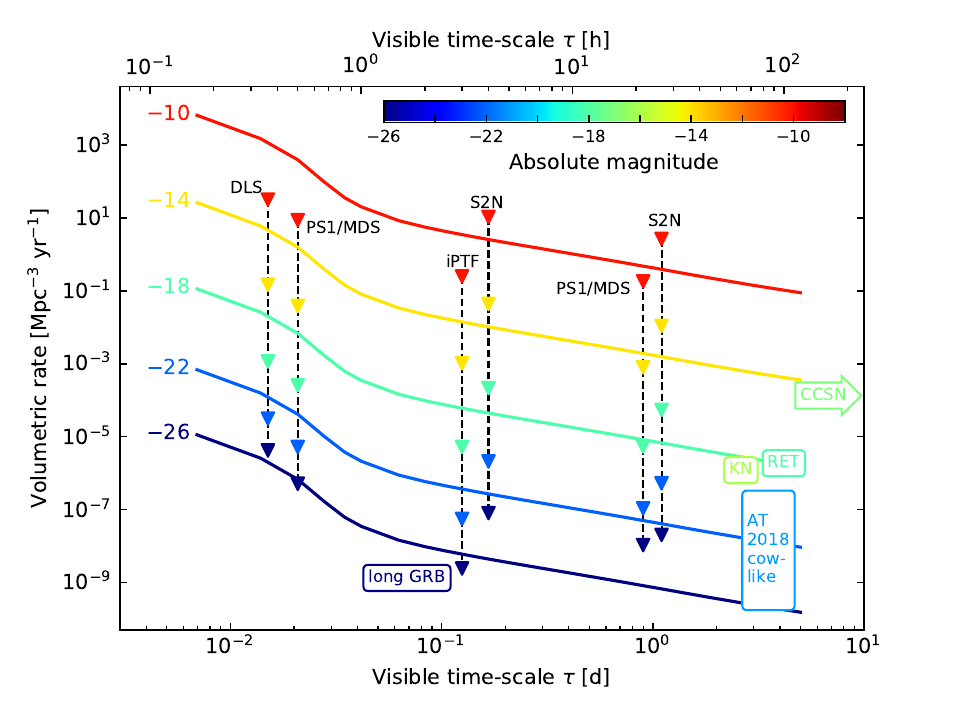}
   \caption{
   \label{fig:Rv}
   Upper limit of the volumetric event rate of extragalactic FOTs as a function of the visible time-scale. 
   The solid curves represent the upper limits from the Tomo-e Gozen survey and triangles represent upper limits from other surveys: 
   DLS \citep{Becker2004}, PS1/MDS \citep{Berger2013}, iPTF \citep{Ho2018}, and S2N \citep{Roestel2019}. 
   The limits for the absolute magnitudes of $-10$ mag, $-14$ mag, $-18$ mag, $-22$ mag, and $-26$ mag are shown from top to bottom (colours according to the colour bar at the top). 
   For comparison, the volumetric event rates of CCSNe ($M\sim -17$ mag, 1.0 $\times \ 10^{-4}$ Mpc$^{-3}$ yr$^{-1}$; \eg \citealt{Li2011,Perley2020}), 
   rapidly evolving transients (RETs, $M\sim -18$ mag, $<1.05 \times 10^{-6}$ Mpc$^{-3}$ yr$^{-1}$; \eg \citealt{Li2023,Pursiainen2018}), 
   KNe ($M\sim -16$ mag, $<9.0 \times \ 10^{-7}$ Mpc$^{-3}$ yr$^{-1}$; \citealt{Andreoni2021}),
   AT 2018cow-like luminous FBOTs
   ($M\sim -21$ mag, $3.0 \times \ 10^{-10}$ to 4.2 $\times \ 10^{-7}$ Mpc$^{-3}$ yr$^{-1}$; \citealt{ho2023}), and 
   long GRBs ($M\sim -26$ mag, 1.0 $\times \ 10^{-9}$ Mpc$^{-3}$ yr$^{-1}$; \citealt{Wanderman2010}, beaming uncorrected rate)
   are shown with their typical time-scale. 
   }
  \end{center}
\end{figure*}

To compare the limits from different surveys with different depths, we discuss the volumetric rate of extragalactic FOTs, $\Rv$, which is the number of extragalactic FOTs per comoving volume per time. 
To evaluate $\Rv$, we need to assume an absolute magnitude of FOTs as the horizon distance depends on the absolute magnitude of FOTs. 
With an assumed absolute magnitude $M$ and a limiting magnitude of the survey, the horizon distance and the comoving volume within the horizon distance $V(M)$ can be obtained. 
Here, $K$ correction is not taken into account 
as the spectral energy distributions of extragalactic FOTs are not certain. Then, the volumetric rate of FOTs is calculated as follows: 
\begin{align}
\Rv \ [\mathrm{Mpc}^{-3}\ \mathrm{yr}^{-1}] =
\frac{\Ra \ [\mathrm{deg}^{-2} \ \mathrm{d}^{-1}]}{V(M) \  [\mathrm{Mpc}^3]}
\nonumber &\\
\times \ \Omega_{\mathrm{sky}} \ [\mathrm{deg}^{2}\ \mathrm{sky}^{-1}] 
\times \ t_{\mathrm{yr}} \ [\mathrm{d}\ \mathrm{yr}^{-1}] ,
\end{align}
where $\Omega_{\mathrm{sky}}$ is the area of the entire sky 
(41,253 deg$^{2}$ sky$^{-1}$) and $t_{\mathrm{yr}} = $ 365.25 d yr$^{-1}$. 
Figure \ref{fig:Rv} shows the upper limits of volumetric rates of the extragalactic FOT from the Tomo-e Gozen survey (solid curves) and other surveys (triangles). 
As expected, for intrinsically luminous objects, i.e., objects that can be detected at greater distances, 
the constraints of the volumetric rate become tighter because of the larger survey volume.

Compared with other surveys, the Tomo-e Gozen survey gives one of the most stringent constraints for intrinsically luminous objects (e.g., absolute magnitude of $< -24 $ mag). 
For deeper surveys such as DLS and PS1, if the objects are intrinsically luminous, the increase of the comoving volume is suppressed at high redshift due to the cosmological effects. 
On the other hand, for wider and shallower surveys such as the Tomo-e Gozen survey, 
the survey volume is dominated by relatively low redshift. 
For example, even for the case of absolute magnitude of $-26$ mag, 
the maximum distance modulus is $\mu$ = 43.8, which corresponds to the redshift of $z \sim 1$. 
As a result, wider/shallower surveys become more effective than deeper/narrower survey as it is less affected by the cosmological effects to the survey volume.

Finally, we discuss the derived upper limits by comparing with the event rates of known transients. 
For example, the local core-collapse SN (CCSN) volumetric rate is  about $10^{-4}$ Mpc$^{-3}$ yr$^{-1}$ \citep{Li2011,mattila2012,Perley2020}. 
This event rate corresponds to the upper limit of FOTs with $M = -18$ mag with a visible time-scale of $\tau \sim 0.08$ d ($\sim 2$ h). 
In other words, FOTs brighter than $-18$ mag with a visible time-scale of $\sim 2$ h should be less frequent than normal CCSNe. 
Also, the observed, beaming uncorrected (on-axis) long-GRB rate is known to be about 
$10^{-9}$ Mpc$^{-3}$ yr$^{-1}$ \citep{Wanderman2010}. 
As mentioned in Section \ref{sec:1}, 
relativistic jetted transients such as GRBs and jetted TDEs are expected to be observed as extragalactic FOTs with an absolute magnitude of $M \leq - 24$ mag on a few hours time-scale. 
For FOTs with $-26$ mag and a visible time-scale of hours, their upper limits are still higher than the rate of on-axis long GRB; therefore, non-detection of on-axis GRB afterglow-like objects with Tomo-e Gozen is consistent with the known event rate of GRBs. 
On the other hand, upper limit derived by iPTF for $-26$ mag on a time-scale of 4 h is comparable with the event rate of the observed GRBs \citep{Ho2018}. 
In fact, \cite{Cenko2015} have discovered GRB afterglow independently from the $\gamma$-ray trigger with iPTF. 
Moreover, there have been several recent reports of discoveries of optical GRB afterglow with ZTF \citep[\eg][]{Andreoni2021,Ho2022}.

For transients with a time-scale of a few days, 
our upper limit for objects with $M = -18$ mag is comparable with event rate of so-called rapidly evolving transients \citep[RET, \eg][]{Pursiainen2018,Li2023}. 
Also, the upper limits for objects of $M < -20$ mag in a time-scale of a few days are within range of the event rate of AT 2018cow-like luminous transients or FBOT estimated by \cite{ho2023}. 
In fact, a recently reported rapid luminous object, AT 2022aedm \citep{nicholl2023}, 
has been detected by the Tomo-e Gozen survey after the searching period in this paper. 
This is consistent with the fact that the upper limits obtained in this study are comparable to the event rate of FBOTs. 

\section{Summary}
\label{sec:5}

In this paper, we present a search for extragalactic FOTs in the Tomo-e Gozen survey. 
Using the data with a cadence of about one hour,
we searched for FOTs with a time-scale of a few hours to a few days. 
By removing Galactic sources using Gaia EDR3, 
we identified 113 candidate objects with detection duration < 5 d. 

Among these 113 candidates, 81 have point source counterparts in the PS1 images. 
These include previously classified objects, eight quasars or blazars and 59 candidates of DNe. 
Ten other objects are associated with galaxies, all of which show typical supernova-like luminosity evolution. 
Hence, they are likely to be normal SNe detected only near the peak. 
The remaining 22 objects are not associated with any counterparts in the PS1 images. 
Their nature is not entirely clear, but a comparison of their light curves combined with ZTF data suggests that many of them are Galactic DNe. 
Therefore, we conclude that no genuine extragalactic FOTs have been discovered in this survey. 

Based on this result, we estimate the upper limits of the volumetric event rate of extragalactic FOTs as a function of their visible time-scale. 
For luminous transients with an absolute magnitude of $-26$ mag, 
the upper limit of the event rate is found as 
4.4 $\times$ $10^{-9}$ Mpc$^{-3}$ yr$^{-1}$ for a time-scale of 4 h, 
and 7.4 $\times$ $10^{-10}$ Mpc$^{-3}$ yr$^{-1}$ for a time-scale of 1 d. 
Thanks to the shallow and wide survey strategy of Tomo-e Gozen, 
which is less affected by the cosmological effects to the survey volume, 
our search provide one of the tightest limits to date on intrinsically luminous extragalactic FOTs with a time-scale of < 1 d.

\section*{Acknowledgements}
We thank Jan van Roestel and Jin Beniyama for fruitful discussion and valuable comments. 
We thank the referee for constructive comments. 
This research was partially supported by the Optical and Infrared Synergetic Telescopes for Education and Research (OISTER) program funded by the MEXT of Japan, JST FOREST Program (grant No. JPMJFR212Y), and the JSPS Grant-in-Aid for Scientific Research (grant Nos. 25103502, 16H02158, 19H00694, 16H06341, 17H06363, 18H01261, 18H05223, 20H00179, 21H04491, 21H04997, 21H04997, 23H00127).

\section*{Data Availability}
The data used in this research will be shared on reasonable request to the corresponding author.




\bibliographystyle{mnras}
\bibliography{reference} 








\bsp	
\label{lastpage}
\end{document}